\documentstyle[12pt]{article}
\newfont{\feff}{cmti10}
\newcommand{\newsection}{    
\setcounter{equation}{0}
\section}
\def\bx{\mbox{\boldmath $x$}}
\def\bw{\mbox{\boldmath $w$}}
\def\bz{\mbox{\boldmath $z$}}

\def\by{\mbox{\boldmath $y$}}
\def\d{{\rm d}}
\def\F{{\rm F}}

\begin{document}
\begin{titlepage}
\rightline{\vbox{\halign{&#\hfil\cr
&UCTP-111/92\cr
&December 1992\cr}}}
\vspace{0.5in}
\begin{center}
{\Large\bf
Hamiltonian Effective Potential}
\medskip
\vskip0.5in

\normalsize Beth Basista and Peter Suranyi
\\ \smallskip
\medskip

{ \sl University of
Cincinnati, Cincinnati, Ohio 45221}\\
\smallskip

\end{center}
\vskip1.0in

\begin{abstract}
A Hamiltonian effective potential (the logarithm of the
square of the wave functional) is defined and calculated at
the tree and one loop levels in a $\phi^4$ scalar field
theory. The loop expansion for eigenfunctionals is equivalent
to the combination of WKB expansion and an expansion around
constant field configurations. The results are compared with
those obtained from the Lagrangian effective potential. While
at tree level the results from the two methods coincide, at
one loop level they differ. Our result at one loop level is
that the reflection symmetry is not broken at $m^2\geq0$.
\end{abstract}\end{titlepage}
\baselineskip=20pt
\newsection{Introduction}

The effective action provides the standard tool for the
investigation of the phase structure of quantum field
theories. The usefulness of the effective action stems from
its extremal property, namely at vanishing external current
its variation with respect to the classical field vanishes.
\begin{equation}
\frac{\delta\Gamma[\phi]}{\delta\phi(x)}=0,
\label{action-principle}
\end{equation}
In other words, classical field configurations minimizing the
effective action are the relevant physical configurations. In
particular, the effective potential, which is the effective
action taken at translation invariant field configurations,
predicts symmetry breaking. If the classical field $\phi$
minimizing the effective potential is nonvanishing and $\phi$
is not symmetric for some of the symmetries of the
Lagrangian, then those symmetries are spontaneously broken.

The widest used systematic method for calculating the
effective potential is the loop expansion~\cite{coleman}. Each
order of the loop expansion sums up infinitely many
perturbative contributions. In fact, the loop expansion is
nothing else but a perturbation expansion at fixed
$z=\sqrt{g/2}(\phi/\mu)$, where $\mu $ is either the physical
mass, or alternatively, the value of $\phi$ at which the
subtraction is performed.  The general form of the expansion
is
\begin{equation}
\Gamma(\phi)= \frac{\mu^4}{g}\sum_{n=0} g^n F_n(z),
\label{loop-expansion}
\end{equation}
where $n$ labels the $n$-loop contribution.

In a recent paper, E. Fradkin~\cite{fradkin} has called
attention to the fact that the ground state wave functional in
Schr\"odinger representation can be used to investigate
the phase structure of quantum field theories, just like
the effective action. Using the ground state wave functional
taken at translation invariant field configurations, the
Hamiltonian effective potential (HEP), $2S^{\rm H}(\phi)$,
can be defined as
\begin{equation}
\left(\Psi_{\rm gr}(\phi)\right)^2\simeq e^{-2S^{\rm
H}(\phi)}.
\label{hamiltonian-action}
\end{equation}

The invariance of matrix elements in transforming from
Heisenberg to Schr\"odinger representation allows one to give
the following expression to the square of the ground state
wave functional:
\begin{eqnarray}
e^{-2S^{\rm H}(\phi)}&=&\int \d J(\bx)
Z[J(\bx)\delta(t)]\exp\{-\int \d^3x J(\bx)
\phi\}\nonumber\\&=&\int \d J(\bx) \langle0|\exp\{\int
\d^3x~J(\bx)\hat\phi(\bx)\}|0\rangle
\exp\{-\int \d^3x~J(\bx)\phi\}
\label{potential-relation}
\end{eqnarray}
Relation (\ref{potential-relation}) was known to
Symanzik~\cite{symanzik}, who proved the renormalizability of
the Schr\"odin-\linebreak ger representation.

The only difference between
the expression given by Symanzik and
(\ref{potential-relation}) is that Symanzik restricted
integration over the field $\chi$ to the half-space $x_0\leq0$.
The equivalence of the definitions will be discussed briefly
in the next section.

The perturbation theory rules one finds for HEP (a slightly
modified version of that of Symanzik) clearly show that HEP
is distinct from the effective action. As an example, we may
point out that one particle reducible diagrams are not fully
canceled in HEP. That makes the Lagrangian, diagrammatic
calculation of the terms of the loop expansion of HEP
difficult. Nevertheless, since HEP is proportional to the
volume, all matrix elements in Schr\"odinger representation are
dominated by values of $\phi$ minimizing HEP. Then HEP is a
useful predictor of the phase structure of the theory as
well.

It is worth mentioning that HEP also differs from the
potential one obtains by the saddle point evaluation of the
right hand side of (\ref{potential-relation}). The relation
of that potential to HEP is similar to that of the action and
the effective action in Lagrangian theories: They differ by
loop terms, even for smooth field configurations.

By using the expression of HEP in terms of the ground
state wave functional, (\ref{hamiltonian-action}), we are able
to obtain a loop expansion for HEP. It turns out that a loop
expansion, similar to (\ref{loop-expansion}), is completely
equivalent to the WKB, or semiclassical expansion, taken at
translational invariant field configurations.

We calculate HEP in zero and one loop orders in a $\phi^4$
theory. The predictions of HEP at tree level for the value of $\langle
\phi\rangle$ are identical to the ones obtained from the tree
level Lagrangian theory in the unbroken and broken
phases. At the one loop level, however, the results differ.
At $m^2=0$ the one loop HEP is unchanged from the tree level
form, keeping the tree level phase structure intact. There is
no symmetry breaking at $m^2>0$ either. This seeming
contradiction with the results obtained using the Lagrangian
effective potential~\cite{coleman}, which predicts symmetry
breaking at $m^2=0$, is resolved if we recall the observation
of Coleman and Weinberg that the obtained phase transition
point is not believable due to the large value of the
coupling constant.

\newsection{ Effective Potentials}

In this section we will define a multitude of potentials.
The minimum of each of these  potentials has some physical
relevance.

Our starting point is the standard field theoretic action,
$S[\phi]$, the minimum of which gives classical field
configurations. The effective potential
formalism~\cite{cheng} defines the effective action
\begin{equation}
\Gamma[\phi]=W[j]-\int dx j(x) \phi(x),
\label{lagrangian-action}
\end{equation}
where $\phi(x)$ is the classical field and $W[j]$ is the
generating functional of connected diagrams. In
(\ref{lagrangian-action}), $j(x)$ should be the solution of
the equation
\begin{equation}
\frac{\delta W[j]}{\delta j(x)}=\phi(x).
\label{classical-field}
\end{equation}
Note that the functional integral over the source gives back
the action
\begin{equation}
\int dj(x) e^{W[j]-\int dx j(x) \phi(x)}\simeq e^{iS[\phi]},
\label{standard-s}
\end{equation}
while the evaluation of the integral on the left hand side of
(\ref{standard-s})
in the saddle point approximation gives the effective action.
These two actions are different even for smooth field
configurations, such as $\phi(x)=\phi=$constant. While they
agree on the tree level, they differ in loop terms.

Alternatively, one can use a source restricted to a spacelike
surface ('Hamiltonian source') in a similar
context~\cite{symanzik}. We choose the surface $t=0$.
Accordingly, let us define the functional $S^{\rm H} [ \phi]$
as follows
\begin{equation}
\int d\chi~e^{iS[\chi]}\delta\left(\phi(\bx)-
\chi(\bx,0)\right)\simeq\int dJ(\bx)
\exp\{W[J(\bx)\delta(t)]-\int d^3x J(\bx) \phi(\bx)\}\simeq
e^{-2S^{\rm H}[\phi]},
\label{hep-definition}
\end{equation}
where $\phi(\bx)$ is a time independent classical field,
while $\chi(\bx,t)$ is a Lagrangian time dependent field. We
define the Hamiltonian effective potential, HEP, as the
functional $2S^{\rm H}[\phi]$, taken at translation invariant
field configurations.

Two comments have to be made concerning
(\ref{hep-definition}). First, Symanzik defined the
Hamiltonian
generating function by restricting the functional integral
over field $\chi$ to the half space $x_0\leq0$. We arrive at
our definition if we take the product of Symanzik's generating
functional with another one for a field theory defined for
$x_0\geq0$. The slight advantage of our definition is that
the momentum space perturbation rules are slightly simpler.

Secondly, if one lifts off regularization, the relation
between the Lagrangian and Hamiltonian fields has to be
modified. Symanzik showed~\cite{symanzik} that a new
renormalization constant, rescaling field $\phi$, must be
introduced
\begin{equation}
Z_5\phi(\bx)=\chi(\bx,0).
\label{rescaling}
\end{equation}

Now (\ref{hep-definition}) can be written as the Fourier
transform of
\[
\langle 0| \exp\{i\int d^3x J(\bx) \hat  \phi(\bx,0)|0\rangle
=
\langle \Psi_{\rm gr}[\phi]|\exp\{i\int d^3x J(\bx)
\phi(\bx)|\Psi_{\rm gr}[\phi]\rangle
\]
where $\hat\phi(\bx,t)$ is the field operator in Heisenberg
representation, $\phi(\bx)$ is the (diagonal) field in
Schr\"odinger representation, and $|0\rangle$ and $|\Psi_{\rm
gr}[\phi]\rangle$ are the ground state in Heisenberg and
Schr\"odinger representations, respectively. Then

(\ref{hep-definition}) states that the HEP is
expressed by the ground state wave functional as
\begin{equation}
e^{-2S^{\rm H}[\phi]}=\Psi_{\rm gr}^2[\phi].
\label{equality}
\end{equation}

One can also define the potential $\Gamma^{\rm H}[\phi]$ as
\begin{equation}
\Gamma^{\rm H }[\phi]=W[J(\bx)\delta(t)]-\int d^3x J(\bx)
\phi(\bx),
\label{new-potential}
\end{equation}
where $J(\bx)$ is the solution of the equation
\begin{equation}
\frac{\delta W[J(\bx)\delta(t)]}{\delta J(\bx)}=\phi(\bx).
\label{hamiltonian-field}
\end{equation}

The relation of  $2S^{\rm H}$ and $\Gamma^{\rm H}$ is
obviously the same as that of the action, $S$, and of the
effective action $\Gamma$: they differ in loop terms but they
are equal at the tree level. We will provide a perturbative
expression for $\Gamma^{\rm H}$ later.

Notice that $\Gamma^{\rm H}(\phi)$ has interesting extremal
properties on its own. It satisfies the equation
\begin{equation}
\frac{\delta\Gamma^{\rm H}[\phi]}{\delta \phi(\bx)}=-J(\bx).
\label{hamiltonian-extremum}
\end{equation}
In other words, $\Gamma^{\rm
H}$ becomes extremal  in the physical, $J(\bx)=0$, case.

Unfortunately, the calculation of
$\Gamma^{\rm H}[\phi]$ is not as simple as that of the
standard effective potential, as we will see below.

Let us now derive diagrammatic expansions for $2S^H$ and
$\Gamma^{\rm H}$.
We introduce a time dependent source for the field
$\chi(\bx,t)$ as well and write (\ref{hep-definition}) as
\begin{eqnarray}
e^{-2S^{\rm H}[\Phi]}&\simeq&\int \d J(\bx) \exp\left\{-i\int
d\bx J(\bx)\Phi(\bx)\right\} \sum_k\frac{1}{k!}\left[ i\int
dx L_{\rm int}(\delta/\delta j(x))\right]^k\nonumber \\
&\times&\int\d\chi
\exp\bigg\{-\frac{1}{2}\int \d x \d y \chi(x)\chi(y)
D^{-1}(x-y)\nonumber \\
&+&i\int \d x~j(x)\chi(x) +i\int \d\bx ~
J(\bx)\chi(\bx,0)\bigg\}\bigg|_{j=0}.
\label{three}
\end{eqnarray}

Now it is a matter of two Gaussian integrations over
$\chi(x)$ and $J(\bx)$ to arrive at the final formula from
which diagrammatic rules can be read off. We obtain
\begin{eqnarray}
e^{-2S^{\rm H}[\Phi]}&\simeq& \exp\left\{-\frac{1}{2}\int
\d\bx \d\by \Phi(\bx)\Phi(\by)\tilde D^{-1}(\bx-\by)\right\}
\sum_k\frac{1}{k!}\left[ i\int dx L_{\rm int}(\delta/\delta
j(x))\right]^k\nonumber \\ &\times&\exp\left.\left\{-
\frac{1}{2}\int \d x \d y j(x)j(y) D_{\rm eff}(x,y)-i\int \d
x \d\by j(x)
\Phi(\by)H(x,\by)\right\}\right|_{j=0}.
\label{four}
\end{eqnarray}

The Green's functions appearing on the right hand side of
(\ref{four}) are defined as follows:
\begin{eqnarray}
D_{\rm eff}(x,y)&=&D_\F(x-y)-\int \d\bz\d\bw D_\F(x-
\bz)\tilde D^{-1}(\bz-\bw)D_\F(\bw-y)\nonumber \\
&=&\theta(x_0y_0)
 [D_F(\bx-\by,x_0-y_0)-D_F(\bx-\by,x_0+y_0)]
\end{eqnarray}
and
\begin{equation}
H(x,\by)=\int \d\bz D_\F(x-\bz)\tilde D^{-1}(\bz-\by).
\label{six}
\end{equation}

Symanzik's effective propagator is exactly the same as
$D_{\rm eff}$ restricted to the half space $x_0\leq0$ and
$y_0\leq0$. Note that $D_\F$ is the usual Feynman propagator,
while $\tilde D(\bx)=D_\F(\bx,t=0)$. Thus, $\tilde D^{-1}(\bx-\by)$
is the inverse of the  $\tilde D$ in {\em three dimensional}
space.

We can read off the the momentum space Euclidean Feynman
rules for the the action $2S[\Phi]$ from (\ref{four}).  For
every internal line one
has
\begin{equation}
D_{\rm eff}(p_0^{(1)},p_0^{(2)},p)=\frac{\delta(p_0^{(1)}-
p_0^{(2)})}{(p_0^{(1)})^2+E(p)^2}-
\frac{\pi}{E(p)[(p_0^{(1)})^2+E(p)^2]
[(p_0^{(2)})^2+E(p)^2]},
\label{seven}
\end{equation}
where $E(p)=\sqrt{m^2+p^2}$. From now on we will mostly deal
with the spatial part of the momentum only, thus, the
notation $p,q,...$ will be used for three momenta.

The diagrammatic rule for the spatial components of the
momentum is the standard one: one has to integrate over loop
momenta and momentum is conserved at vertices.
There is a delta function at every vertex for the
conservation for the 0th component of mometum vectors, but
one has to integrate over the 0th component of all of the
lines emanating from the vertex. An external $\phi$ field
requires a multiplier of
\begin{equation}
\phi\frac{m}{\pi}\frac{1}{p_0^2+m^2},
\label{eight}
\end{equation}
where $p_0$ is the 0th component of the momentum carried by
$\phi$ from a vertex.

The Feynman rules for $\Gamma^H$ are similar, except inside

loops
not propagators of the form (\ref{seven}), but rather
standard Feynman propagators, must be used. Then loop
contributions to $2S^H$ and $\Gamma^H$ are obviously
different, but at the tree level the two potentials agree.

We have used the above Feynman rules to derive the first few
terms of the tree level effective potential:
$2S^H(\phi)\simeq\Gamma^{\rm H}(\phi)$
\begin{equation}
2S^H(\phi)= m\phi^2+ \frac{g\phi^4}{48m}-
\frac{g^2\phi^6}{4608m^3}
+O(\phi^8)+{\rm loop~terms}.
\label{first-terms}
\end{equation}

Note that $2S^H(\phi)=\Gamma^{\rm H}(\phi)$ is obviously
different from the standard effective potential because the
sixth order tree diagram term, coming solely from
one--particle reducible diagrams, is not completely canceled.
Thus we have established that $2S^{\rm H}$ and
$\Gamma^{\rm H}$ are distinct. They both differ from the
effective action or the field theoretic action, as well. The
calculation of these effective potentials in the Lagrangian
formalism is complicated. On the other hand, as we will see
later, a systematic approximation scheme, based on the WKB
approximation and  analogous to the loop expansion, can be
performed in Schr\"odinger representation. Among others, we
will be able to check (\ref{hamiltonian-action}) directly by
comparing (\ref{first-terms}) with a similar expression
obtained there.

 \newsection{Wave functionals in the tree approximation}

We turn to the discussion of the Schr\"odinger representation
now~\cite{jackiw}. The eigenfunctionals satisfy the time
independent Schr\"odinger equation \[
H |\Psi[\phi]\rangle=E|\Psi[\phi]\rangle.
\]
In a $\phi^4$ field theory,
 the Hamiltonian has the form
 \begin{equation}
 H={1\over2}\int \d^3x \left(-
\frac{\delta^2}{\delta\phi(x)^2}+m^2\phi(x)^2+\partial_\mu
\phi(x)\partial_\mu\phi(x)+  \frac{g}{12}\phi^4(x)\right).
\label{hamiltonian}
 \end{equation}

 We can use fields in momentum representation as well. Using
the relation  \begin{equation}
 \phi(p)=\frac{1}{(2\pi)^{3/2}}\int
\d^3x\phi(x)e^{ip\cdot\bx},  \label{fourier}
 \end{equation}
 we obtain the following form for the Hamiltonian
 \begin{equation}
 H={1\over2}\int \d^3p \left[-
\frac{\delta^2}{\delta\phi(p)\delta\phi(-
p)}+(m^2+p^2)\phi(p)\phi(-p)\right]+
\frac{g}{4!(2\pi)^3}\int
\prod_1^4[\phi(p_i)\d^3p_i]\delta^3(\sum_j^4p_j) .
\label{hamiltoniantwo}
 \end{equation}
\subsection{Wave functional for the ground state and
spontaneous symmetry breaking} The ground state wave
functional is sought in the form  \begin{equation}
\Psi_{\rm GR}[\phi]=\exp\{-S[\phi]\}
\label{ansatz}
\end{equation}
because it has no nodes~\cite{feynman}. Notice that we have
dropped the superscript $H$ from the notation of HEP. This
can be done with impunity, since from now on we will deal
with Hamiltonian theories only.

Substituting (\ref{ansatz}) into the Schr\"odinger equation,
we obtain the following functional differential
equation~\cite{cornwall}:
  \begin{eqnarray}
E^{(0)}&=&{1\over2}\int \d^3p\left[\frac{\delta^2
S}{\delta\phi(p)\delta\phi(-p)}-\frac{\delta
S}{\delta\phi(p)}\frac{\delta S}{\delta \phi(-p)}+(m^2+p^2)
\phi(p)\phi(p)\right]\nonumber\\&+&\frac{g}{4!(2\pi)^3}
\int\prod_{i=1}^4 \d^3p_i\phi(p_i)~\delta(\sum p_i).
\label{s-equation}
\end{eqnarray}

The term of (\ref{s-equation}) containing the second
functional derivative of $S$ generates loop diagrams. In a
zeroth order WKB approximation, that term is dropped since
$S$ is supposed to be proportional to $1/\hbar$. Then only
tree diagrams are generated. These are in general momentum
dependent, due to the presence of the kinetic term,
proportional to $p^2$ in (\ref{s-equation}). In a tree
diagram, if all external momenta are set equal to zero, then
all internal momenta are zero as well. Thus, omitting the
kinetic term generates tree diagrams at $p=0$.

It is easy to solve (\ref{s-equation}) for $S$ when the
second derivative and kinetic terms  are omitted. Then for
translation invariant $\phi$, defined by the choice of the
function $\phi(p)$ as
\begin{equation}
 \phi(p)=(2\pi)^{3/2}\delta^3(p)\phi,
 \label{invari-phi}
 \end{equation}
 it reduces to the ordinary differential equation
\begin{equation}
U(\phi)=\left(\frac{1}{V}\frac{\d
S^0}{\d\phi}\right)^2=m^2
\phi^2+\frac{g}{12}\phi^4-\frac{2E^{(0)}}{V}.
\label{constant-phi}
\end{equation}
In what follows, we will drop the volume factors from in front
of $S$ and $E$ and these quantities will be understood as
densities.

We will argue that (\ref{constant-phi}) determines not only
$S^0$, but also the ground state energy, $E^{(0)}$. A
physical solution should satisfy the following requirements:
\begin{itemize}
\item The derivative of $S$ is single valued and real
for all $\phi$. If $S$ was complex, $\Psi_{\rm GR}$ would not
be unique. Furthermore it would have nodes, what is forbidden
for a ground state wave functional~\cite{feynman}.
\item The derivative $\d S/\d \phi$ is an odd
 function of $\phi$. In other words, $S(\phi)$, and

 consequently $\Psi_{\rm GR}(\phi)$, are even. Otherwise
 $\Psi_{\rm GR}(\phi)-\Psi_{\rm GR}(-\phi)$ would be a
 ground state wave functional as well and it would have a node
 at $\phi=0$. That is not allowed, as it has been pointed out
 above.

\item $S$ has an absolute minimum for some $\phi$. This is, of
course, a necessary condition for having a normalizable wave
functional. This condition implies that
$\d S/\d \phi\rightarrow\pm\infty$
if $\phi\rightarrow\pm\infty.$
\end{itemize}

The listed conditions determine $E^{(0)}$ uniquely. They
imply that $E^{(0)}$ should be set to make the point(s) where
the absolute
minimum is attained to be a zero(s) of $U(\phi)$ as well.
Oddness and boundedness of  $\d S/\d \phi$
requires then that we have to
switch
the choice of the sign of the root taken from $U(\phi)$ in
(\ref{constant-phi}) at $\phi=0$.

At $m^2>0$ the minimum of $U(\phi)$ is at $\phi=0$ and
thus $E^0=0$ at the tree level. We obtain
\begin{equation}
\frac{d S^0}{d\phi}=m\phi\sqrt{1+\frac{g\phi^2}{12m^2}}.
\label{ds-not-broken}
\end{equation}

If  $m^2<0$  the minimum is attained at $\phi=\pm
|m|\sqrt{6/g}$.
Then one has to choose $E^{(0)}=-3m^4/(2g)$ and
\begin{equation}
\frac{d S^0}{d\phi}=\epsilon(\phi)
\sqrt{\frac{g}{12}}\left(\phi^2-
\frac{6|m|^2}{g}\right), \label{ds-broken}
\end{equation}

$S^0$ is obtained from (\ref{ds-not-broken})
and  (\ref{ds-broken}) as
\begin{equation}
S^0=\frac{4m^3}{g}\left[\left(1+
\frac{g\phi^2}{12m^2}\right)^{3/2}-1\right], \label{solution}
\end{equation}
for $m^2>0$, while it is
\begin{equation}
  S^0=
\frac{\sqrt{g}}{3\sqrt{12}}\left[|\phi|^3-
\frac{18|m|^2}{g}|\phi|\right]  \label{breaking-solution}
\end{equation}
for $m^2<0$.  $S^0$ for $m^2=0$ can
be obtained either from (\ref{solution})
or from (\ref{breaking-solution}) since they go over into each
other continuously.

Note that the HEP's (\ref{solution}) and (\ref{breaking-solution})
are quite different from the usual
tree level effective potentials obtained in Lagrangian theory.
When (\ref{solution}) is expanded in $\phi^2$ it
agrees with (\ref{first-terms}), the calculation of HEP
performed in the framework of the Lagrangian
formalism. At  the critical point, $m=0$, this function has the
behavior $S\sim g^{1/2 }\phi^3$, unlike the Lagrangian
effective potential in the tree approximation, which has the
behavior $g\phi^4$.  The expectation value of $\phi$ is to be
determined by the minimum of $S^0$, owing to the large
multiplier $V$ in the exponent. In other words, using

according to our current notations
\[\Psi{\rm GR}(\phi)=\exp\{-VS(\phi)\}.\]
The minima of $S^0$ are at 0 for
$m^2>0$ and at $\phi_0=\pm|m|\sqrt{6/g}$, exactly at the same
points as in the Lagrangian theory. Furthermore, the
second derivatives of $S^0$, the effective masses,

at the minimum point also
coincide. However, away from the minimum the the Lagrangian
and Hamiltonian theories are quite
different.

\subsection{ Small oscillations around the constant field
configuration}

Defining the deviation of the field $\phi(x)$ from the
constant value $\phi$ as $\chi(x)$, and the Fourier
transform of $\chi(x)$ as $\chi(p)$, we have the following
expansion \begin{equation}
S[\phi]=S(\phi)+\frac{1}{2}\int
\d^3p\chi(p)\chi(p)T_{p}(\phi)+O(\chi^3),
\label{small-oscillation}
\end{equation}
where the function $T_{p}(\phi)$ is defined as
\begin{equation}
\delta^3(p_1+p_2)T_{p}(\phi)=\left.\frac{
\delta^2S[\phi]}{\delta\phi(p_1)\delta
\phi(p_2)}\right|_{\phi(x)=\phi}. \label{t-def}
\end{equation}

For the discussion of excited states and also for the
calculation of loop corrections, one needs the second
functional derivative of the tree level effective potential
taken at a translation invariant field value.

One can easily obtain an equation for $T_{k}(\phi)$ if one
differentiates the equation for

 $S^0[\phi]$ twice with respect to $\phi(k)$ and $\phi(-k)$.
One obtains
\begin{eqnarray}
0&=&\int \d^3p \bigg[-\frac{\delta^3S^0}{\delta\phi(k)\delta
\phi(-k)\delta\phi(p)}\frac{\delta S^0}{\delta\phi(-p)} -
\frac{\delta^2 S^0}{\delta\phi(p)\delta\phi(k)}
\frac{\delta^2 S^0}{\delta\phi(-p)\delta\phi(-k)}\nonumber\\
&+&\frac{g}{2(2\pi)^3} \phi(p) \phi(-p) \bigg]
+\frac{V}{(2\pi)^3}(m^2+k^2)=0.
\label{differentiate}
\end{eqnarray}
Setting $\phi(p)=(2\pi)^{3/2}\phi\delta^3(p)$, we obtain the
following differential equation for $T$
\begin{equation}
{T'}_{k}(\phi)\frac{\d
S^0}{\d\phi}+T_{k}(\phi)^2-m^2-k^2-\frac{g}{2} \phi^2=0,
\label{t-equation}
\end{equation}
where the prime denotes a derivative with respect to $\phi$.

 Differential equation (\ref{t-equation}) can be easily
solved. Its solution, regular at $\phi=0$, is
\begin{equation}
{T}_{k}(\phi)= \left[\omega+3m^2
z^2\frac{2m\sqrt{1+z^2}+\omega}{2m^2+\omega^2+3z^2m^2+
3\omega m\sqrt{1+z^2}}\right], \label{mom-dep-solution}
\end{equation}
where we use $\omega=\sqrt{k^2+m^2}$ and
$z=\sqrt{g/12}(\phi/m)$.

For $m^2<0$ one can also solve differential equation

(\ref{t-equation}) easily. The solution regular at
$\phi=\phi_0=m\sqrt{6/g}$, or equivalently at $z=1/\sqrt{2}$,
positive definite (as
required by (\ref{small-oscillation}) ), and an even function
of $\phi$ is
\begin{equation}
{T}_{k}(\phi)=\omega+3|m|^2(2z^2-1)
\frac{2|m||z|+\omega}{2\omega^2-|m|^2+6|m|^2z^2+6|m||z|\omega},
\label{broken-solution}
\end{equation}
where $z$ is defined as before  but
$\omega=\sqrt{2|m|^2+k^2}$.

At $m^2=0$ we define $z=\sqrt{g/12}\phi$. Then the form of
$T_{k}(\phi)$ could be obtained either from
(\ref{mom-dep-solution}) or from (\ref{broken-solution}). We
obtain
\begin{equation}
{T}_{k}(\phi)= k+3z^2\frac{2|z|+k}{k^2+3z^2+3k|z|}.
\label{zero-mass-solution}
\end{equation}

Note that in the tree approximation, when $\phi$ is taken at
the minimum of $S(\phi)$, $T_{k}(\phi)$ reduces to
$\sqrt{k^2+\mu^2}$ where $\mu$ is the physical mass. This
agrees with the expression for the corresponding density
function (\ref{small-oscillation}) for the free theory, taken
at the modified physical mass.

 \subsection{Wave functional for the first excited state}
We can write the wave functional for the first excited state
as
\begin{equation}
  \Psi_{\rm EXC}[\phi]=\chi[\phi,q]\Psi_{\rm GR}[\phi].
\label{excited}
  \end{equation}
  The functional $\chi$ satisfies the following functional
differential equation~\cite{suranyi}   \begin{equation}
\int \d^3p\left[\frac{\delta
\chi[\phi,q]}{\delta\phi(p)}\frac{\delta S}{\delta
\phi(-p)}-\frac{1}{2}\frac{\delta^2
\chi[\phi,q]}{\delta\phi(p)\delta\phi(-
p)}\right]=E(q)\chi[\phi,q], \label{excited-equation}
\end{equation}
where $E(q)$ is the excitation energy. $\chi[\phi,q]$ is an
odd functional in the unbroken phase. Again, in the leading
order of the WKB approximation, the term containing the second
derivative is dropped (the term kept contains only one multiplier
of $1/\hbar$). After setting the field
at a translational invariant value and taking $q=0$, we
obtain the following differential equation for $\chi$
\begin{equation}
\chi'(\phi)S'(\phi)=E(0)\chi(\phi),
\label{chi-equation}
\end{equation}
which is solved at $m^2>0$ as
\begin{equation}
\chi(\phi)=\exp\left\{\frac{E(0)}{m}\int
\frac{\d\phi}{\phi\sqrt{1+\frac{g\phi^2}{12m^2}}}\right\}
=\left(\frac{z+\sqrt{z^2+1}-1}{z+\sqrt{z^2+1}+1}
,
\right)^{E(0)/m}, \label{chi-zero}
\end{equation}
where we use the notation $z=(\phi/m)\sqrt{g/12}$.
This solution is analytic at $\phi=0$ only if $E(0)=nm$,
where $n$ is an integer. The lowest energy state is obtained
at $n=1$. In other words, the physical mass is $m$. Then we
have
\begin{equation}
\chi(\phi)=\frac{z+\sqrt{z^2+1}-1}{z+\sqrt{z^2+1}+1}
\label{chi-solution}
\end{equation}
and $\chi$ is an odd function of $\phi$ as a first excited state
should be.

Alternatively, we can take the functional derivative of
(\ref{excited-equation}) with respect to $\phi(k)$, then set
$\phi$ to its translational invariant value
(\ref{invari-phi}). All terms will be proportional to
$\delta^3(q-k)$ from momentum conservation. Then
$\eta(\phi,q)$, defined by
\begin{equation}
\eta(\phi,q)\delta^3(k-q) =
{\delta\chi[\phi,q]\over\delta\phi(k)}\bigg|_{\phi(x)=\phi}
\label{excited-function}
\end{equation}
satisfies the following differential equation
\begin{equation}
\eta'(\phi,q){S}'+\eta(\phi,q)T_q(\phi)
=E(q)\eta(\phi,q),
\label{Chi-equation}
\end{equation}
where $T_q(\phi)$ has been defined in
(\ref{mom-dep-solution}).

The solution of differential equation (\ref{Chi-equation})
has the form
\begin{equation}
\eta(\phi,q)=\exp\left\{\int \d\phi\frac{E(q)-
T_q(\phi)}{S'(\phi)}
\right\}.
\label{excited-solution}
\end{equation}
Now observe that after scaling out the mass, ${S^0}'$ is
proportional to $\phi$ at low $\phi$. Since $T_q(\phi)
=\omega+O(\phi^2)$, $\eta(\phi,q)$ is regular at $\phi=0$
only if $E(q)=\sqrt{q^2+m^2}+nm$, where $n$ is an integer.
The lowest energy state of these states is the first excited
state, with energy $E(q)=\sqrt{q^2+m^2}$. In other words, the
regularity of the excited state wave functional fixes the the
excitation spectrum.

 The solution of (\ref{chi-equation}) in the broken phase is
\begin{equation}
\chi(\phi)=\exp\left\{\pm\frac{E(0)}{\sqrt{g/12}}\int
\frac{\epsilon(\phi)d\phi}{\phi^2+6m^2/g}\right\}=
\left[\frac{\sqrt{2}|z|-1}{\sqrt{2}|z|+1}
\right]^{E(0)/(\sqrt{2}m}. \label{chi-zero-new}
\end{equation}
$\chi(\phi)$ is analytic around the point $\phi=\pm\phi_0$
($z=\pm1/\sqrt{2}$) if we choose $E(0)=n\sqrt{2}|m|$.

The lowest
energy state is again obtained at $n=1$. The physical mass is
$\sqrt{2}|m|$ in agreement with the Lagrangian result. One
has then \begin{equation}
\chi(\phi)=\frac{\sqrt{2}|z|-1}{\sqrt{2}|z|+1}
. \label{broken-chi-solution}
\end{equation}
The energy eigenvalues at nonvanishing momentum can be
analyzed in a manner similar to that of the unbroken phase.

Notice that
there is no need to shift fields to arrive at the correct
physical mass in the broken phase similar to
 the effective potential formalism.

 \subsection{Normalization
and orthogonality of wave functionals}

Note that the only
important contribution to the wave functionals comes from the
saddle point, the minimum of $S$. Expanding around the minima
we can see that the normalized wave functionals in the
unbroken phase are \begin{eqnarray}
\Psi_{\rm GR}&\simeq &V^{1/4}\exp\left\{-
\frac{Vm\phi^2}{3}\right\}[1+O(V\phi^4)], \label{norm-gr}\\
\Psi_{\rm EXC}&\simeq &V^{3/4}\phi\exp\left\{-
\frac{Vm\phi^2}{3}\right\}[1+O(\phi^2)+O(V\phi^4)].
\label{norm-exc}
\end{eqnarray}
The orthogonality of these is already assured by the fact
that $\chi(\phi)$ is odd, while  $S(\phi)$ is even.

In the broken phase the normalized wave functionals have the
form  \begin{eqnarray}
\Psi_{\rm GR}&\simeq &V^{1/4}\exp\left\{-\frac{Vm(\phi-
\phi_0)^2}{\sqrt{2}}\right\}[1+O(V(\phi-\phi_0)^3)],
\label{norm-gr-broken}\\
\Psi_{\rm EXC}&\simeq &V^{3/4}(\phi-\phi_0)
\exp\left\{\frac{-Vm(\phi-\phi_0)^2}{\sqrt{2}}
\right\}[1+O(\phi-\phi_0)+ O(V(\phi-\phi_0)^3)].
\label{norm-exc-broken}
\end{eqnarray}
The overlap of the ground and excited state wave functions is
of $O(V^{-1/2})$, vanishing in the large volume limit.

\newsection{Wave functional in the one loop approximation and
renormalization}
In the previous section, we have determined HEP in the tree
approximation. Now we set out to find the one loop correction
to
HEP. To accomplish that task we have to eliminate divergences
and
renormalize HEP.

Symanzik proved the renormalizability of field theories in
Schr\"odinger representation~\cite{symanzik}. In one loop
order, the only modifications of  (\ref{s-equation}) come from
the appearance of the contribution of the second functional
derivative and from the necessity of renormalization. We
obtain
\begin{equation}
Z_5^{-
2}\left(\frac{\d S}{\d\phi}\right)^2=
\frac{1}{(2\pi)^{3-\epsilon}} \int \d^{3-\epsilon}k
{}~T_{k}(\phi)+Z_2Z_5^2m^2\phi^2+Z_1Z_5^4\frac{g}{12}\phi^4-
2E, \label{new-equation}
\end{equation}
where $T_{k}(\phi)$ is the
second functional derivative of $S^0$ with respect to the
field
$\phi(p)$ as given  in (\ref{mom-dep-solution}),
(\ref{broken-solution}), and (\ref{zero-mass-solution}) and
taken at translational invariant field configuration. $Z_1$
is the coupling renormalization constant, $Z_2$ is the
multiplicative mass renormalization constant, and $Z_5 $ is
the new renormalization constant defined by (\ref{rescaling}) and
 introduced by
Symanzik~\cite{symanzik}, whose
notations we are using.  Note that no wave function
renormalization is needed in one loop order. Writing these
renormalization constants as \begin{equation}
Z_i=1+a_i\frac{g}{32\pi^2\epsilon}+O(g^2),
\label{renorm_const}
\end{equation}
where $\epsilon=3-D$, we have $a_1=6$, $a_2=-2$, and $a_5=-1$.
The renormalization constants get modified by {\em finite}
terms of $O(g)$, which depend on the renormalization scale.
The resulting freedom will be resolved by using appropriate
initial conditions for HEP.

Notice that in multiplying (\ref{new-equation}) by $Z_5^2$ the
renormalization constants appear
in the combination $Z_1Z_5^6$ in the interaction term.
 This combination is finite in
one loop order as shown by (\ref{renorm_const}).
Consequently, one expects no divergent term of the form
$\delta g\phi^4/24$ to appear when divergences are removed
from the integral of $T_{k}(\phi)$.

The momentum integral over function $T_k(\phi)$, appearing in
(\ref{new-equation}), is divergent when $D\rightarrow3$. One
obtains three potentially divergent terms only; namely,
\begin{equation}
T_{k}(\phi)=T_{k}^{\rm reg}(\phi)+t_0+a \phi^2+
2b\phi\frac{\d S^0}{\d\phi} ,
\label{subtraction}
\end{equation}
where superficially the integrals over $t_0$, $a$, and $b$ are

respectively quartically,
quadratically, and linearly divergent. For an appropriate choice

of the constants
$t_0$, $a$, and $b$, the integral  over $T_{k}^{\rm
reg}(\phi)$ is convergent.

The three terms removed from $T$ have the following physical
meaning. The constant term contributes to the one loop value
of the ground state energy.
The second term contributes to mass
renormalization. Finally, the third term could be
incorporated  as an addition to
$\d S^0/\d\phi$ of a term proportional to
$\phi$ combined with a further mass renormalization.
In other words, this term would necessitate a subtraction
in HEP of the form $\delta\mu \phi^2$. Indeed, Symanzik has
shown that a linearly divergent counterterm $\delta\mu\int
\d^3x \phi^2(x)/2$ must be included in the perturbative
expression for the effective action. It was pointed out by
L\"uscher~\cite{luscher}, however, that there is no need to
introduce an extra infinite renormalization constant

if one uses dimensional
regularization. Indeed, the integral over coefficient $b$ of
(\ref{b-def}) is finite if we define it by continuing in the
number of dimensions below two, calculating the convergent
integral and then continuing back to the physical dimension,
three. Since we have  no physically plausible way of fixing
the freedom arising from such a subtraction, we will
insist on using
dimensional regularization.

Our contentions concerning the regularization of the integral
over $T_k(\phi)$ can be made transparent by expanding
$T_k(\phi)$ in an asymptotic series of $1/\omega$.

The following two expansions are obtained for the $m^2>0$
and $m^2<0$ cases, respectively:
\begin{equation}
T_k(\phi)=\omega+3z^2\frac{m^2}{\omega}-3z^2
\sqrt{1+z^2}\frac{m^3}{\omega^2}+3z^2
\frac{m^4}{\omega^3}+O(1/\omega^4)
\label{asymptotic}
\end{equation}
and
\begin{equation}
T_k(\phi)=\omega+3(z^2-1/2)\frac{|m|^2}{\omega}
-3|z|(z^2-1/2)\frac{|m|^3}{\omega^2}
+3(z^2-1/2)\frac{|m|^4}{\omega^3}+O(1/\omega^4),
\label{asymptotic-two}
\end{equation}
where as before $z^2=g\phi^2/(12|m|^2)$.
Note that while in (\ref{asymptotic}) $\omega^2=m^2+k^2$, in
(\ref{asymptotic-two}) $\omega^2=2|m|^2+k^2$.

The momentum integral over $1/\omega^2$ is
convergent in the dimensional
renormalization scheme. Then in both of the equations
(\ref{asymptotic}) and (\ref{asymptotic-two}), the only
divergent terms left are either constant or they are
proportional  to the square of the field. The coefficient of
$1/\omega^2$ is proportional to $\phi \d
S/\d\phi$, as it has been remarked earlier.

As far as finite renormalizations are concerned, besides the
energy and the mass, charge is also renormalized.
The values of the three subtraction constants must be set by
conditions imposed on $U(\phi)=(\d
S/\d\phi)^2$. One could demand that
the
second and fourth derivatives of $U(\phi)$ are equal to
predetermined constants at some $\phi=\phi_0$. We
will use a simpler, alternative scheme
in which
 \begin{itemize}
\item The mass renormalization constant will be fixed by
demanding
\begin{equation}
\left.\frac{\partial U(\phi)}{\partial(\phi^2)}
\right|_{\phi=\phi_0}=m^2+\frac{1}{6}g\phi_0^2,
\end{equation}
where $m^2$ is the mass parameter appearing in the
renormalized Lagrangian and where $g$ is the renormalized
charge.
\item  The finite charge renormalization
constant will be fixed by the condition
\[
\left.\frac{\partial^2U(\phi)}{(\partial
\phi^2)^2}\right|_{\phi=\phi_0}=\frac{g}{6}.
\]
\item The constant subtraction will be fixed by the
requirement that $\d S/\d \phi\rightarrow
\pm\infty$ for $\phi\rightarrow\pm\infty$ and that

$\d S/\d \phi$ is continuous at its absolute minimum.

As it has been
explained in the discussion of the tree level potential, this
implies that  $U(\phi)$ vanishes at the point(s) where it
attains its absolute minimum. \end{itemize}

 \subsection{Renormalization of the ground state functional
for $m^2>0$}
We will subtract at the physical point $\phi=0$.
Then explicit expressions for the
coefficients $a$, $b$, and $t_0$ of (\ref{subtraction}) are
\begin{eqnarray}
t_0&=&\omega,\label{tnot-def}\\
b&= &-\frac{gm(4m+\omega)}{8(m+
\omega)^2(2m+\omega)},
\label{b-def}\\ a&=&\frac{g}{4(m+\omega)}-2b.
\label{a-def}\\
\end{eqnarray}
As before, we use the notation $\omega=\sqrt{m^2+k^2}$.
Using the renormalization procedure described in the
previous section
we can
easily integrate $T_k(\phi)$ over the

$(3-\epsilon)$-momentum
to obtain the following expression for $U(\phi)$

at $m^2>0$.
In the limit of $\epsilon\rightarrow0$
\begin{equation}
U(\phi)=\frac{12m^4}{g}[U_0(z)+gU_1(z)]-2E^{(0)}
+\frac{1}{(2\pi)^3}\int \d^3p\omega(p),
\label{final-form}
\end{equation}
where, as before $z^2=g\phi^2/(12m^2)$, and where the zero
and one loop contributions to $U(\phi)$, $U_0(z)$ and
$U_1(z)$ have the following form
\begin{equation}
U_0(z)=z^2+z^4,
\label{u0}
\end{equation}
and
\begin{eqnarray}
U_1(z)&=&\frac{z^2}{8\pi^2}\bigg\{
\frac{w_1\sqrt{w_1^2-1}(2\sqrt{1+z^2}-w_1)}{\sqrt{1-3z^2}}
\log(\sqrt{w_1^2-1}+w_1)
\nonumber \\
&-&(w_1\rightarrow w_2)
+cz^2-12(\sqrt{1+z^2}-1)\bigg\}.
\label{u1}
\end{eqnarray}
The constant $c$ is given by
\[ c=6+4\sqrt{3}\log(2+\sqrt{3}), \]
and the functions $w_1$ and $w_2$ are defined as
\[w_{1,2}=\frac{3}{2}\sqrt{1+z^2}\mp\frac{1}{2}
\sqrt{1-3z^2}.\] Notice
that according to the above defined renormalization
prescription,  $U_1(z)\sim z^6$ at small $z$.

The behavior of
$U_1(\phi) $ can be analyzed either analytically or
numerically.  It is a positive, monotonically increasing
function. Then at $g\leq0$\ \ $U(\phi)$ is monotonically
increasing as well.  The absolute minimum of $U(\phi)$ is at
$\phi=0$, consequently one has to set  \[E^{(0)}
=\frac{1}{2(2\pi)^3}\int \d^3p\omega(p). \]
Thus, at every $g$, the behavior of $U(\phi)$ near the
absolute  minimum, $z=0$,  is dominated by the tree level
potential. In other  words, the symmetry is not broken at any
$g$ for $m^2>0$.
\subsection{Renormalization at $m^2=0$}
 The discussion of the $m^2=0$ case is even simpler.
 Explicit expressions for the
coefficients $a$, $b$, and $t_0$ of (\ref{subtraction}),  are
then
\begin{eqnarray}
t_0&=&|k|,\label{tnot-def-two}\\
b&= &0,
\label{b-def-two}\\ a&=&\frac{g}{4|k|}.
\label{a-def-two}\\
\end{eqnarray}

Due to the fact that there is no divergent term
proportional to $\phi^4$ in the integral of $T$ over momentum,
there is no infrared divergence even at vanishing mass. Thus
in Schr\"odinger representation, one can subtract at
$\phi=0$ even in the massless case. Then the integral of
$T_k(\phi)$ over
$p$ is easy to perform and we obtain
\begin{equation}
U(\phi)=\frac{(g+\delta g)}{12}\phi^4
+\frac{g^2}{16\pi\sqrt{3}}\phi^4,
\end{equation}
where $\delta g$ is the finite renormalization of the
coupling  constant. No matter where we subtract, $\delta
g$ should be chosen to cancel the last term of $U(\phi)$. In
other words,  the form of $U(\phi)$, thus that of $S(\phi)$,
is exactly the same at the one loop level as at the tree
level. The reflection  symmetry, $\phi\rightarrow-\phi$  is
not
broken at $m^2=0$. This result contradicts the one
obtained from the Lagrangian effective
potential~\cite{coleman}. As it has been emphasized, however,
that result uses couplings and field values outside the
bounds of a perturbative expansion,  and, as such, it is
meaningless~\cite{coleman}.

\subsection{Renormalization at $m^2<0$}

Using the freedom of choice of the subtraction point, it is
most convenient to choose  $\phi_0^2=6|m^2|/g$, or $z^2=1/2$,
the `physical
point', as the subtraction point. By this choice, the
physical mass is set at $\sqrt{2}|m|$.

The divergences of the integral over (\ref{broken-solution})
are removed and $T_k^{\rm reg}(\phi)$ is made to vanish at
$z^2=1/2$ as $(z^2-1/2)^3$ by subtracting the following
three terms from $T_k(\phi)$:
\begin{equation} T^{\rm reg}_q(\phi)=T_q(\phi)-t_0-
a(2z^2-1)-b(2z^2-1)^2,
\label{subtraction-broken}
\end{equation}
where
\begin{eqnarray}
t_0&=&w\label{tnot-broken}\\
a&=&
\frac{3|m|^2}{2w+\sqrt{2}|m|}
\label{a-broken}\\
b&=& -\frac{3|m|^3(w+2\sqrt{2}|m|)}{\sqrt{2}(2w+ \sqrt{2}|m|.
)^2(w+\sqrt{2}|m|)}\label{b-broken}
\end{eqnarray}

After the constant subtraction term has been absorbed by the
ground state energy, we obtain the potential $U(\phi)$ in the
following form:
\begin{equation}
U(\phi)=\frac{12|m|^4}{g}[U_0(z)+gU_1(z)],
\label{u-form}
\end{equation}
where
\begin{equation}
U_0(z)=\left(z^2-\frac{1}{2}\right)^2,
\label{u-zero}
\end{equation}
and
\begin{eqnarray}
U_1(z)&=&\frac{3(2z^2-1)}{4\pi^2}\bigg[\frac{w_1
\sqrt{w_1^2-2}(2|z|-w_1)}{\sqrt{2-3z^2}}\log
\frac{\sqrt{w_1^2-2}+w_1}{2}
\nonumber \\ &-&(w_1\rightarrow w_2)
+\frac{\pi}{2\sqrt{3}}-\frac{2z^2-1}{4}\left(
\frac{4\pi}{\sqrt{3}}-3\right)\bigg].
\label{u-one}
\end{eqnarray}
The quantities $w_{1}$ and $w_2$ are defined as
\begin{equation}
w_{1,2}=\frac{3}{2}|z|\mp\frac{1}{2}\sqrt{2-3z^2}.
\label{ws}
\end{equation}

Then $U(\phi)$ defined by (\ref{u-form}) has absolute minima
at $z=\pm1/\sqrt{2}$ for all $g$ satisfying
\[ g<g_0=\frac{16\pi^2\sqrt{3}}{2\pi-3\sqrt{3}}\simeq 256.\]
For $g>g_0$,\ $U(\phi)$ becomes unbounded from below. Of
course, such a large value of $g$ is outside the bounds of
the loop expansion.
\subsection{Renormalization of the excited state wave
functional}

The functional $\chi[\phi]$, appearing in the excited state
wave functional satisfies (\ref{excited-equation}). At one
loop level and at translation invariant field configuration
we obtain the following equation:
\begin{equation}
Z_5^{-2}\frac{\d
\chi(\phi,q)}{\d\phi}\frac{\d S}{\d
\phi}-\frac{1}{2(2\pi)^{3-\epsilon}
}\int d^{3-\epsilon}k~U_k(\phi,q)=E(q)\chi(\phi,q),
\label{excited-one-loop}
\end{equation}
where we defined
\begin{equation}
U_k(\phi,q)=\left.\frac{\delta^2
\chi[\phi,q]}{\delta\phi(p)\delta\phi(-
p)}\right|_{\phi(x)=\phi}. \label{second-derivative}
\end{equation}

Notice that, unlike at the renormalization of the ground state,
the field renormalization is not canceled, thus, a divergence,
proportional to $\chi'S'$ must appear at the integration over
$U_k(\phi,q)$. To show this, it is sufficient to find the
asymptotic expansion of $U_k(\phi,q)$ at
$k\rightarrow\infty$.

A differential equation for $U_k(\phi,q)$ is obtained by

functionally differentiating (\ref{excited-equation}) twice with
respect to the field, after dropping the second functional
derivative term. The equation has the form
\begin{equation}
{U'}_k(\phi,q)S'+2U_k(\phi,q)T_k(\phi)+\chi'(\phi,q){T'}_k
(\phi)=E(q)U_k(\phi,q).
\label{u_equation}
\end{equation}
Though (\ref{u_equation}) can be solved exactly in terms of
quadratures, it is more instructive to solve it
asymptotically, for large $k$.

Iterating (\ref{u_equation}) leads to
\begin{eqnarray}
U&=&-\frac{1}{2T-E}\left[\chi'T'-S'\frac{\d}{\d\phi}
\left(\frac{1}{2T-E}\chi'T'\right)\right]+
O(1/k^4)\nonumber \\&=&-\frac{\phi\chi'}{4\omega^2}
+\frac{S'\chi'}{4\omega^3}+O(1/k^4)
\label{expanded-u}
\end{eqnarray}
where for the sake of simplicity
we have dropped arguments and subscripts and where we have
used the asymptotic expansion
\[ T_k=\omega+\frac{\phi^2}{4\omega}-\frac{\phi S'}
{4\omega^2}+O(1/k^3).\]

The dimensionally regularized integral over the first
term of the asymptotic expansion of $U$ is convergent.
The logarithmically divergent second term of

(\ref{expanded-u}) gives
\begin{equation}
\frac{1}{2(2\pi)^{3-\epsilon}}\int

\d^{3-\epsilon}k~U_k(\phi,q)=
\frac{\chi' S' }{ 16 \pi^2 \epsilon}+({\rm terms~finite~at~}
\epsilon=0),
\end{equation}
precisely canceling the divergent part of the first term of
(\ref{excited-one-loop}), $Z_5^{-2}\chi'S'$. Thus all
divergences from the equation for the excited state wave
functional are removed by the renormalization constant $Z_5$.

\newsection{Conclusions}

We have shown in this paper that the ground state wave
functional in Schr\"odinger representation allows for the
investigation of the phases of field theories just as
expeditiously as the effective potential. We have calculated
tree level and one loop contributions, but
the calculation of higher loop contributions is also feasible
in
the HEP formalism. In fact, higher loop contributions can be
written down in closed form, in terms of quadratures.

Though the introduction of HEP has a theoretical interest on
its own, the real physical interest lies in the investigation
of theories with gauge fields and with fermions. Hamiltonian
theories containing these fields can be written down
with equal ease~\cite{jackiw}.
\section*{Acknowledgements}

The authors gratefully acknowledge the support by the U.S.

Department of Energy under Grant
number

DE-FG02-ER84-40153. One of the authors (P.S.) also thanks
the Theory Group at the Stanford Linear Accelerator Center
for its hospitality.


\begin{thebibliography}{10}
\small
\addtolength{\itemsep}{-6pt}

\bibitem{coleman}
S. Coleman and E. Weinberg,

{\sl Phys. Rev.}
{\bf  D7}, 1888 (1973);
\bibitem{fradkin} E. Fradkin, ``Wave Functionals for Field
Theories and Path Integrals'', Illinois preprint P-92-4-48;
\bibitem{symanzik} K. Symanzik, {\sl Nucl. Phys.} {\bf B190
[FS3]}, 1 (1981);
\bibitem{jackiw}
R. Jackiw, ``Analysis on Infinite-Dimensional
Manifolds---Schr\"odinger Representation for Quantized
Fields'', Seminaire de Math\'ematiques Sup\'erieures,
Montr\'eal, Qu\'ebec, Canada, June 1988.
\bibitem{feynman}
R. Feynman, {\sl Nucl. Phys.} {\bf B188}, 479 (1981);
\bibitem{cheng}
Ta-Pei Cheng and Ling-Fong Li, {\sl Gauge Theory

of Elementary Particle Physics}, Clarendon, Oxford 1984;
\bibitem{luscher}

M. L\"uscher, {\sl Nucl. Phys.} {\bf B254}, 52 (1985);
\bibitem{cornwall}
J. M. Cornwall, {\sl Phys. Rev.} {\bf  D38}, 656 (1988);
\bibitem{suranyi}
P. Suranyi, { \sl Modern Phys. Lett.} {\bf 7}, 1975 (1992).

\end{thebibliography}
 \end{document}